\documentclass[letterpaper]{article}
\usepackage{aaai2027}
\usepackage[hyphens]{url}
\usepackage{graphicx}
\usepackage{natbib}
\usepackage{caption}
\usepackage{subcaption}
\usepackage{algorithm}
\usepackage{algorithmic}

\usepackage{newfloat}
\usepackage{listings}
\DeclareCaptionStyle{ruled}{labelfont=normalfont,labelsep=colon,strut=off}
\floatstyle{ruled}
\newfloat{listing}{tb}{lst}{}
\floatname{listing}{Listing}

\usepackage{booktabs}
\usepackage{tabularx}
\usepackage{amsmath}
\usepackage{enumitem}
\usepackage{longtable}

\usepackage[hidelinks]{hyperref}

\title{Is Bash All You Need? \\An Empirical Study of Tool Interfaces for Enterprise Digital Worker Agents}
\author{
    Hazel Mak\textsuperscript{\rm 1}\thanks{Corresponding author: hazelmak@microsoft.com},
    Susheel Suresh\textsuperscript{\rm 1},
    Sahil Bhatnagar\textsuperscript{\rm 1},\\
    Barry Wang\textsuperscript{\rm 2},
    Chhaya Methani\textsuperscript{\rm 1},
    Alejandro Gutierrez Munoz\textsuperscript{\rm 1}
}
\affiliations{
    \textsuperscript{\rm 1}Microsoft Corporation\\
    One Microsoft Way, Redmond, WA 98052, USA\\
    \textsuperscript{\rm 2}Carnegie Mellon University\\
    5000 Forbes Avenue, Pittsburgh, PA 15213, USA
}

\begin{document}
\maketitle

\begin{abstract}
    In this study, we examine whether a general shell can outperform specialized tools on enterprise tasks. Shell-based agents have shown strong results in coding, but enterprise work also involves moving between applications and services, coordinating with coworkers, and performing professional analysis. We compare five tool interfaces on TheAgentCompany and APEX-Agents using Opus-4.8 and GPT-5.5: typed tools, typed tools plus bash, bash alone, bash with persistent agent-synthesized tools, and programmatic tool calling (PTC), which runs programs whose actions are restricted to a typed tool catalog. Bash alone outperforms typed tools on both benchmarks, improving score by 21.8--24.5 pp on TheAgentCompany and 4.8--7.4 pp on APEX-Agents while using 19--72\% fewer total tokens. Adding typed tools or persistent tool synthesis to bash produces no detectable pooled score gain. PTC uses fewer tokens than direct typed calls with broadly similar task performance, but generally underperforms bash alone in both quality and cost efficiency. For enterprise practitioners, these results favor bash alone when arbitrary execution can be isolated and PTC when security or compliance policies require a fixed tool catalog.

\end{abstract}

\section{Introduction}
\label{sec:intro}
    Can a shell outperform specialized tools on enterprise agent tasks? Frontier models are trained to use bash as a native action language. In coding, mini-SWE-agent exposes only bash~\cite{bib:minisweagent}, while Live-SWE-agent topped SWE-bench Verified by adding agent-created tools~\cite{bib:livesweagent}.

Enterprise products are adopting both catalog-based and shell-based orchestration. Anthropic and OpenAI offer programmatic tool calling over typed catalogs through their APIs~\cite{bib:anthropic_ptc,bib:openai_ptc}. Shell-based offerings include Anthropic's Claude Code~\cite{bib:claudecode}, OpenAI's Codex CLI and SDK~\cite{bib:codex_platform}, and Microsoft's Copilot Studio, which uses the GitHub Copilot CLI and SDK~\cite{bib:copilotstudio_ghharness}.

However, controlled comparisons of shell execution and typed tool interfaces on enterprise workflows remain limited, leaving practitioners with little evidence to guide interface selection.

We conduct a controlled tool-interface ablation on two enterprise workplace benchmarks: \textbf{TheAgentCompany}~\cite{bib:agentcompany}, spanning software company roles, and \textbf{APEX-Agents}~\cite{bib:apexagents}, covering investment banking, management consulting, and corporate law. We compare five interfaces: \textsc{Tool-only}, \textsc{Bash+Tool}, \textsc{Bash}, \textsc{Bash+Synthesis}, and programmatic tool calling (\textsc{PTC})~\cite{bib:anthropic_ptc,bib:openai_ptc}.

Across both benchmarks and models, \textsc{Bash} outscores both shell-free interfaces while remaining among the cheapest options. Adding typed or synthesized tools to \textsc{Bash} yields no detectable pooled score gain. \textsc{PTC} uses fewer tokens than \textsc{Tool-only} but underperforms \textsc{Bash} in task score and token efficiency.

Our contributions are twofold: (i) a controlled comparison of tool interfaces spanning direct tool calling and code-based orchestration; and (ii) an evaluation of their quality and cost on realistic enterprise workflows across diverse professional roles.

\section{Related Work}
\label{sec:related}
    \begin{figure*}[!t]
\centering
\includegraphics[width=0.95\textwidth]{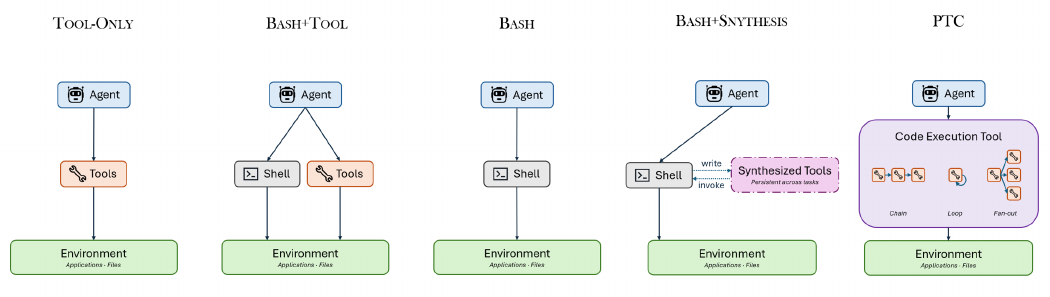}
\caption{Five interfaces formed from shell execution, typed tools, persistent synthesized tools, and restricted programs over typed tools.}
\label{fig:conditions}
\end{figure*}

\begin{figure*}[!t]
\centering
\begin{subfigure}[t]{0.48\textwidth}
\centering
\includegraphics[width=\linewidth]{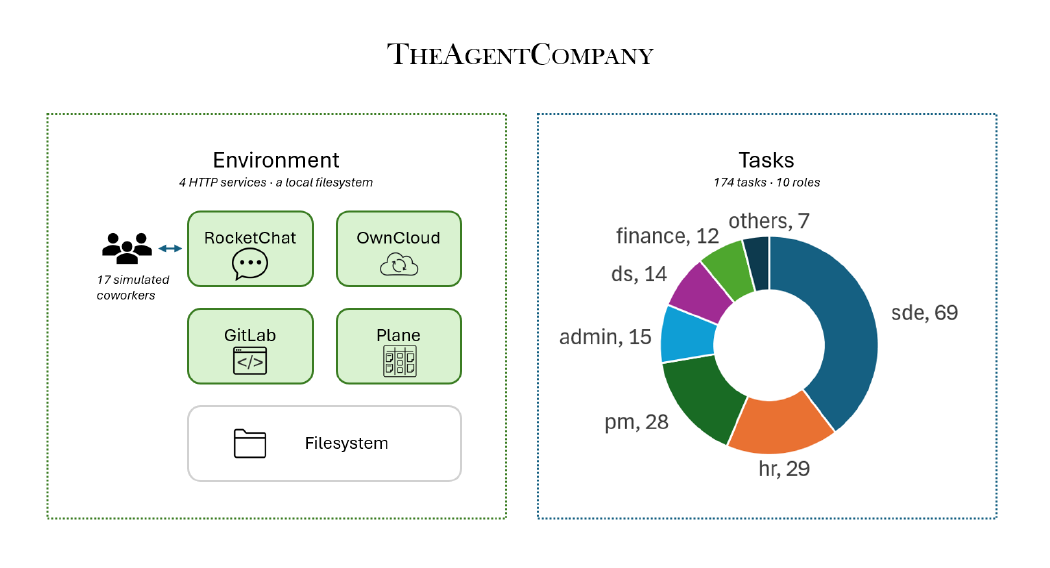}
\caption{TheAgentCompany: 4 self-hosted services, a local workspace, and 17 simulated coworkers reachable through RocketChat.}
\label{fig:environment_tac}
\end{subfigure}
\hfill
\begin{subfigure}[t]{0.48\textwidth}
\centering
\includegraphics[width=\linewidth]{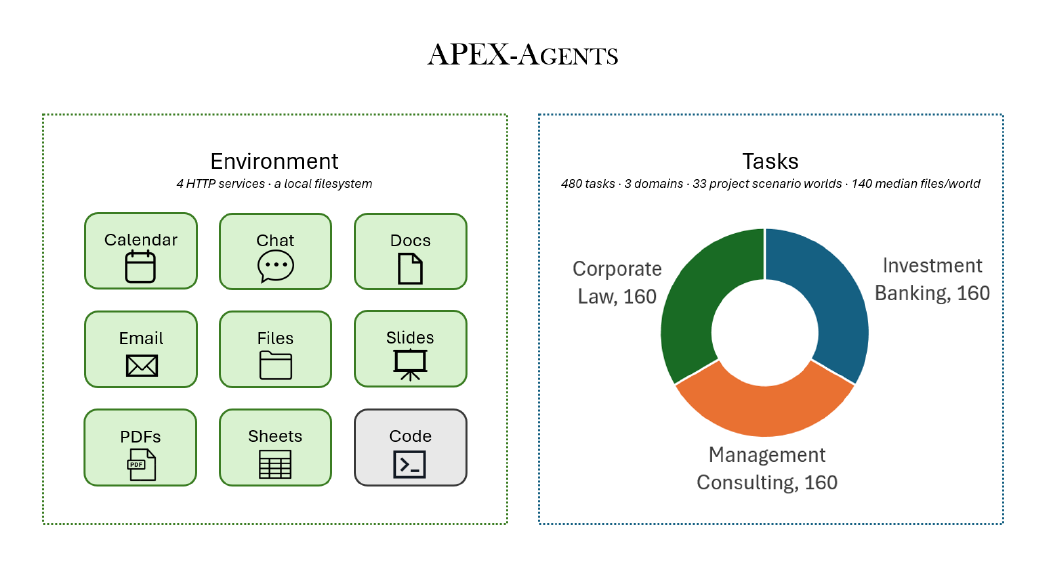}
\caption{APEX-Agents: 9 MCP servers over a shared workspace, with tasks organized into 33 scenario worlds.}
\label{fig:environment_apex}
\end{subfigure}
\caption{Benchmark environments and task types.}
\label{fig:environments}
\end{figure*}

\paragraph{From tool calls to executable actions.}
Typed tool calling gives agents access to external functions~\cite{bib:react,bib:toolformer,bib:gorilla}, with function-call capability evaluated by BFCL~\cite{bib:bfcl}. Code-based approaches extend this interaction by expressing reasoning~\cite{bib:pal,bib:pot}, actions~\cite{bib:codeact,bib:openhands}, and control flow~\cite{bib:llmcompiler,bib:rah,bib:llmascode} in executable programs. Programmatic tool calling applies this idea to a fixed catalog of tools~\cite{bib:anthropic_ptc,bib:openai_ptc}. Together, these approaches motivate our comparison of direct calls, catalog-constrained programs, and shell execution.

\paragraph{Shell-based agents and tool synthesis.}
Coding agents illustrate the range of interface choices: specialized tools in SWE-agent~\cite{bib:sweagent} on SWE-bench~\cite{bib:swebench,bib:swebenchverified}, shell-only execution in mini-SWE-agent~\cite{bib:minisweagent}, self-authored tools in Live-SWE-agent~\cite{bib:livesweagent}, and a fixed workflow in Agentless~\cite{bib:agentless}. Tool synthesis and persistent skill libraries further allow agents to package capabilities for reuse~\cite{bib:toolmaker,bib:creator,bib:voyager,bib:craft,bib:trove,bib:hasp}, including extending available catalogs~\cite{bib:atlass} and supporting computer control~\cite{bib:friday}. These designs motivate testing whether typed tools or persistent synthesized tools add value to a shell-based enterprise agent.

\paragraph{Evaluation beyond function calling.}
Enterprise benchmarks evaluate agents on customer service~\cite{bib:taubench}, web applications~\cite{bib:workarena}, desktop tasks~\cite{bib:osworld}, and workplace workflows~\cite{bib:agentcompany,bib:apexagents}. Concurrently, \citet{bib:bitterlesson_tool} compares programmatic and native JSON tool calling on a BFCL v4 subset. Our independent study instead compares shell execution, direct tool calling, and programmatic tool calling on realistic enterprise workflows.

\section{Method}
\label{sec:method}
    An agent's \emph{tool interface} defines how it can act on its environment, shaping its capabilities, deployment safety, and operating cost. A shell permits arbitrary code execution, allowing many actions in one step but requiring controls on execution and access. A curated catalog instead exposes only deployer-authorized functions, with overhead when called individually. Programmatic tool calling~\cite{bib:anthropic_ptc,bib:openai_ptc} combines these calls in a program that can chain, loop over, and parallelize them while restricting environment actions to the catalog.

We compare five tool interfaces (Figure~\ref{fig:conditions}) with the same model, task text, base prompt, and stopping criteria, adding interface-specific prompts for synthesis and PTC. \textsc{Bash} exposes only the shell, \textsc{Tool-only} exposes only the typed catalog, and \textsc{Bash+Tool} gives the agent both the shell and the catalog. \textsc{Bash+Synthesis} starts with shell access and a persistent directory. Tasks run sequentially, with prompts encouraging the agent to create reusable tools and invoke them on later tasks to reduce discovery costs. \textsc{PTC} lets the model write a restricted program over the typed catalog. Our implementation uses Python as the program language and exposes the catalog as ordinary functions within a restricted runtime that provides no shell or other action channel.

\section{Experiment Setup}
\label{sec:exp_setup}
    \begin{table*}[!t]
\centering
\small
\setlength{\tabcolsep}{4pt}
\begin{tabular}{@{}l l l l r r r r r r@{}}
\toprule
\textbf{Interface} & \textbf{Model} & \textbf{Score (95\% CI)} & \textbf{Pass rate (95\% CI)} & \textbf{Input} & \textbf{Output} & \textbf{Cache-read} & \textbf{Total} & \textbf{\$/task} & \textbf{Time/task} \\
\midrule
\multicolumn{10}{@{}l}{\emph{TheAgentCompany}} \\
\addlinespace[1pt]
\textsc{Tool-only} & Opus-4.8 & 44.9\% {\scriptsize [38.7\%, 51.0\%]} & 24.1\% {\scriptsize [17.8\%, 30.5\%]} & 929k & 16k & 855k & 945k & \$1.20 & 5.1m \\
 & GPT-5.5 & 45.3\% {\scriptsize [39.3\%, 51.3\%]} & 25.3\% {\scriptsize [19.0\%, 31.6\%]} & 458k & 7k & 408k & 465k & \$0.66 & 1.8m \\
\addlinespace[2pt]
\textsc{Bash+Tool} & Opus-4.8 & 67.2\% {\scriptsize [61.3\%, 73.0\%]} & 50.0\% {\scriptsize [42.5\%, 57.5\%]} & 563k & 6k & 537k & 569k & \$0.55 & 3.1m \\
 & GPT-5.5 & 66.3\% {\scriptsize [60.3\%, 72.2\%]} & 47.7\% {\scriptsize [40.2\%, 55.2\%]} & 503k & 6k & 462k & 508k & \$0.60 & 2.5m \\
\addlinespace[2pt]
\textsc{Bash} & Opus-4.8 & 69.4\% {\scriptsize [63.7\%, 74.9\%]} & 50.6\% {\scriptsize [43.1\%, 58.0\%]} & 258k & 6k & 239k & 264k & \$0.37 & 3.0m \\
 & GPT-5.5 & 67.1\% {\scriptsize [61.4\%, 72.7\%]} & 47.7\% {\scriptsize [40.2\%, 55.2\%]} & 370k & 7k & 331k & 377k & \$0.56 & 2.6m \\
\addlinespace[2pt]
\textsc{Bash+Synthesis} & Opus-4.8 & 67.1\% {\scriptsize [61.2\%, 72.7\%]} & 47.7\% {\scriptsize [40.2\%, 55.2\%]} & 244k & 5k & 224k & 249k & \$0.35 & 3.1m \\
 & GPT-5.5 & 66.4\% {\scriptsize [61.0\%, 71.8\%]} & 45.4\% {\scriptsize [37.9\%, 52.9\%]} & 429k & 7k & 388k & 436k & \$0.61 & 3.0m \\
\addlinespace[2pt]
\textsc{PTC} & Opus-4.8 & 52.9\% {\scriptsize [46.8\%, 59.1\%]} & 29.9\% {\scriptsize [23.0\%, 36.8\%]} & 395k & 12k & 362k & 406k & \$0.64 & 3.2m \\
 & GPT-5.5 & 45.1\% {\scriptsize [39.2\%, 51.2\%]} & 24.1\% {\scriptsize [17.8\%, 30.5\%]} & 232k & 10k & 207k & 242k & \$0.54 & 1.9m \\
\addlinespace[2pt]
\midrule
\multicolumn{10}{@{}l}{\emph{APEX-Agents}} \\
\addlinespace[1pt]
\textsc{Tool-only} & Opus-4.8 & 41.1\% {\scriptsize [37.4\%, 44.8\%]} & 24.2\% {\scriptsize [20.4\%, 28.1\%]} & 384k & 13k & 338k & 397k & \$0.73 & 3.4m \\
 & GPT-5.5 & 39.8\% {\scriptsize [36.1\%, 43.4\%]} & 23.3\% {\scriptsize [19.6\%, 27.1\%]} & 322k & 6k & 282k & 328k & \$0.52 & 2.1m \\
\addlinespace[2pt]
\textsc{Bash+Tool} & Opus-4.8 & 47.2\% {\scriptsize [43.5\%, 50.8\%]} & 29.8\% {\scriptsize [25.6\%, 34.0\%]} & 386k & 10k & 350k & 396k & \$0.60 & 3.0m \\
 & GPT-5.5 & 45.5\% {\scriptsize [41.8\%, 49.3\%]} & 29.6\% {\scriptsize [25.6\%, 33.5\%]} & 253k & 5k & 218k & 257k & \$0.42 & 2.0m \\
\addlinespace[2pt]
\textsc{Bash} & Opus-4.8 & 48.5\% {\scriptsize [44.9\%, 52.2\%]} & 30.0\% {\scriptsize [26.0\%, 34.2\%]} & 205k & 10k & 177k & 215k & \$0.48 & 3.5m \\
 & GPT-5.5 & 44.6\% {\scriptsize [40.8\%, 48.3\%]} & 28.3\% {\scriptsize [24.2\%, 32.5\%]} & 171k & 5k & 137k & 176k & \$0.40 & 2.4m \\
\addlinespace[2pt]
\textsc{Bash+Synthesis} & Opus-4.8 & 46.9\% {\scriptsize [43.3\%, 50.7\%]} & 28.3\% {\scriptsize [24.4\%, 32.5\%]} & 226k & 11k & 201k & 237k & \$0.50 & 3.7m \\
 & GPT-5.5 & 45.9\% {\scriptsize [42.2\%, 49.7\%]} & 29.2\% {\scriptsize [25.2\%, 33.1\%]} & 222k & 5k & 186k & 228k & \$0.44 & 6.6m \\
\addlinespace[2pt]
\textsc{PTC} & Opus-4.8 & 40.1\% {\scriptsize [36.5\%, 43.7\%]} & 22.3\% {\scriptsize [18.5\%, 26.2\%]} & 324k & 12k & 294k & 337k & \$0.61 & 3.4m \\
 & GPT-5.5 & 37.8\% {\scriptsize [34.2\%, 41.5\%]} & 20.6\% {\scriptsize [17.1\%, 24.4\%]} & 251k & 5k & 207k & 256k & \$0.48 & 2.1m \\
\addlinespace[2pt]
\bottomrule
\end{tabular}
\caption{\emph{Score}: effort-weighted checkpoint rate (TheAgentCompany) or mean rubric score (APEX-Agents); \emph{pass rate}: fully solved fraction; both with 95\% bootstrap CIs. Token columns: means/task in thousands; input includes cache-read; total = input + output. \emph{\$/task}: mean estimated inference cost at GitHub Copilot list rates; \emph{time/task}: mean wall-clock minutes.}
\label{tab:ablation_summary}
\end{table*}

\begin{figure*}[t]
\centering
\includegraphics[width=0.95\textwidth]{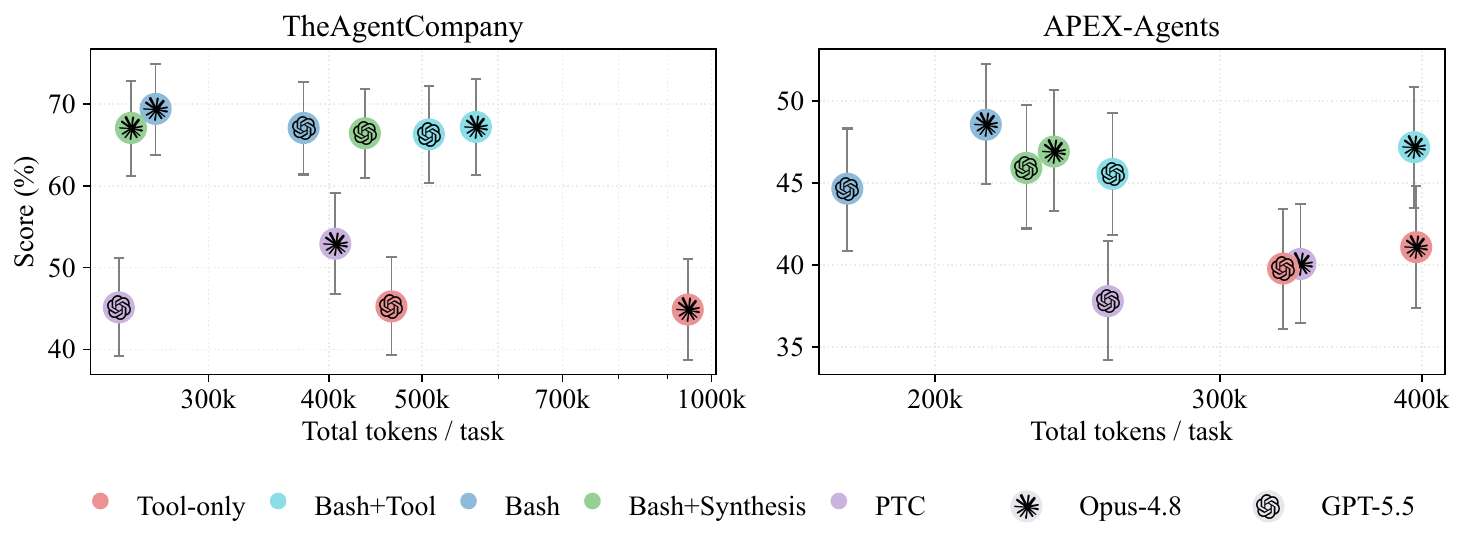}
\caption{Score (\%) versus total tokens/task (log scale). Logos: models; fills: interfaces; vertical bars: 95\% bootstrap CIs. Higher-left is better.}
\label{fig:quality_cost}
\end{figure*}

We run evaluations on TheAgentCompany and APEX-Agents under all five interfaces with \textbf{Opus-4.8} and \textbf{GPT-5.5} at default reasoning levels.

\subsection{TheAgentCompany}
\label{sec:exp_tac}

TheAgentCompany~\cite{bib:agentcompany} simulates a software company where agents use shared services, work with local files, and collaborate with simulated coworkers to complete coding and office workflows (Figure~\ref{fig:environment_tac}).

The benchmark provides no LLM-facing typed tool catalog. We construct a task-oriented catalog of 60 typed tools covering service APIs and workspace operations across the 174 evaluated tasks. The catalog includes 54 service tools and six workspace file tools (Appendix~\ref{app:exp_tac}).

TheAgentCompany coworkers are simulated by GPT-5. We patch coworker reliability defects in the benchmark release: missed messages, premature termination, and stale state. The fixes apply uniformly across interfaces. Grading uses the benchmark's Python checkpoint evaluators.

\subsection{APEX-Agents}
\label{sec:exp_apex}

APEX-Agents~\cite{bib:apexagents} evaluates professional analysis in investment banking, management consulting, and corporate law. Agents use shared-workspace files to answer questions or create and edit documents, spreadsheets, and presentations (Figure~\ref{fig:environment_apex}).

We evaluate all 480 tasks using a fixed June 2026 Archipelago snapshot with nine MCP servers, rather than the later eleven-server version, to avoid reliance on paid financial-data APIs. We retain its 20 typed application tools and use its code-execution server as \textsc{Bash}: a guardrailed Unix shell over the workspace (Appendix~\ref{app:exp_apex}).

An Archipelago-derived LLM judge scores rubric criteria using the final response and workspace-file changes.

\section{Results}
\label{sec:results}
    \subsection{Core Outcomes}
\label{sec:results_core}

We compare task quality, token use, and inference cost across interfaces (Table~\ref{tab:ablation_summary}; Figure~\ref{fig:quality_cost}). \textsc{Bash} outscores both shell-free interfaces in every benchmark-model arm while remaining among the cheapest options. Compared with \textsc{Tool-only}, it uses fewer tokens and lowers estimated inference cost throughout.

\textsc{PTC} consistently saves tokens over \textsc{Tool-only}, but improves score only on TheAgentCompany with Opus-4.8; elsewhere scores are similar or slightly lower. It scores below \textsc{Bash} throughout, with a token advantage only on TheAgentCompany with GPT-5.5.

Adding typed or synthesized tools to \textsc{Bash} generally increases token use without a detectable pooled score improvement in paired comparisons (Appendix~\ref{app:pairwise}). The sole token-use exception is \textsc{Bash+Synthesis} on TheAgentCompany with Opus-4.8.

Benchmark differences are clearest in \textsc{Bash}'s larger pass-rate gains over \textsc{Tool-only} on TheAgentCompany, despite similar baselines on both benchmarks. Flexible execution may help more with coding and cross-application workflows than with domain reasoning, consistent with the domain breakdown and failure analysis (Appendices~\ref{app:by_slice} and~\ref{app:failure_taxonomy}).

Model differences suggest workload-specific choices. On TheAgentCompany, Opus-4.8 scores higher with fewer tokens when using the shell without a typed catalog, favoring it for similar coding and workflow tasks; its \textsc{PTC} score advantage requires more tokens. GPT-5.5 is a token-efficient choice for direct typed-tool access, achieving similar scores on both benchmarks. Document-heavy professional analysis presents a tradeoff: across all APEX-Agents interfaces, Opus-4.8 scores higher while GPT-5.5 uses fewer tokens.

\subsection{Tool Use}
\label{sec:results_tool_use}

We compare number of tool calls, total tool-call tokens, tool call success rate, and exact repetition across interfaces (Table~\ref{tab:tool_use}). \textsc{Bash} reduces number of tool calls and tool-call tokens (more than halved on TheAgentCompany) relative to \textsc{Tool-only}, with similar call-success rates.

\begin{table}[!h]
\centering
\small
\setlength{\tabcolsep}{2pt}
\begin{tabularx}{\columnwidth}{@{}>{\raggedright\arraybackslash}X r r r r@{}}
\toprule
\textbf{Interface} & \shortstack[r]{\textbf{\# Tool calls}\\\textbf{/task}} & \shortstack[r]{\textbf{Tool tokens}\\\textbf{/task}} & \textbf{Success} & \textbf{Repeated} \\
\midrule
\multicolumn{5}{@{}l}{\emph{TheAgentCompany}} \\
\addlinespace[1pt]
\textsc{Tool-only} & 14.0 & 19.6k & 93.3\% & 16.2\% \\
\textsc{Bash+\allowbreak Tool} & 13.5 & 11.3k & 91.5\% & 4.0\% \\
\textsc{Bash} & 12.8 & 9.0k & 92.1\% & 0.3\% \\
\textsc{Bash+\allowbreak Synthesis} & 14.2 & 9.7k & 92.1\% & 0.3\% \\
\textsc{PTC} & 8.8 & 8.4k & 65.2\% & 3.8\% \\
\midrule
\multicolumn{5}{@{}l}{\emph{APEX-Agents}} \\
\addlinespace[1pt]
\textsc{Tool-only} & 13.0 & 15.6k & 90.9\% & 1.7\% \\
\textsc{Bash+\allowbreak Tool} & 11.5 & 13.8k & 90.7\% & 0.4\% \\
\textsc{Bash} & 7.5 & 10.4k & 91.1\% & 0.0\% \\
\textsc{Bash+\allowbreak Synthesis} & 9.5 & 12.1k & 87.7\% & 0.2\% \\
\textsc{PTC} & 9.5 & 12.8k & 86.2\% & 0.2\% \\
\bottomrule
\end{tabularx}
\caption{Tool use statistics. \emph{\# Tool calls/task}: median tool calls/task. \emph{Tool tokens/task}: median tool-name, argument, and result tokens/task; excludes context/cache effects and is not billed usage. Internal Bash/PTC operations excluded; only returned results count. \emph{Success}: calls reporting success; \emph{repeated}: exact tool-and-argument repeats within a task.}
\label{tab:tool_use}
\end{table}

Adding typed tools or synthesis to \textsc{Bash} increases calls and tool-call tokens without consistently improving call reliability.

\textsc{PTC} has the lowest tool-call success rate on both benchmarks. This may reflect the added difficulty of coordinating tool calls and handling intermediate failures within generated programs.

\textsc{Tool-only} has the highest exact-repeat rate on both benchmarks. This may be because shell-based and programmatic interfaces can bundle repeated operations into one invocation, whereas \textsc{Tool-only} exposes each repetition as a separate tool call.

\subsection{Paired Performance by Task Complexity}
\label{sec:results_call_count}

\begin{figure*}[!t]
\centering
\includegraphics[width=0.95\textwidth]{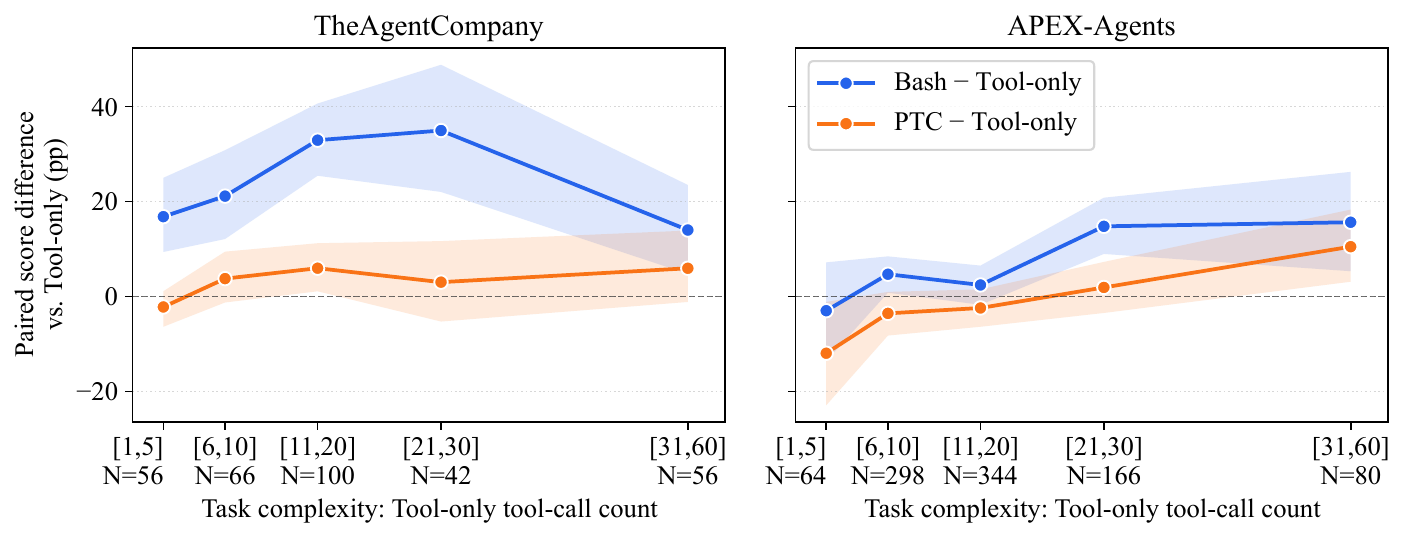}
\caption{Mean paired score differences (pp) from \textsc{Tool-only}: \textsc{Bash} (blue), \textsc{PTC} (orange). Bins: mean \textsc{Tool-only} calls across models; bands: 95\% bootstrap CIs; $N$: task-model pairs/bin. Budget-clipped counts are lower bounds.}
\label{fig:score_vs_tool_count}
\end{figure*}

We examine how \textsc{Bash} and \textsc{PTC} score gains over \textsc{Tool-only} vary with task complexity (Figure~\ref{fig:score_vs_tool_count}). We use task-model paired score differences, binning tasks by their two-model average \textsc{Tool-only} call count as a complexity proxy.

\textsc{Bash}'s score gains are larger on TheAgentCompany than on APEX-Agents. On TheAgentCompany, its advantage peaks at intermediate counts, while \textsc{PTC} shows smaller, uneven gains. On APEX-Agents, both interfaces gain more at high counts, where \textsc{PTC}'s mean score difference becomes positive and \textsc{Bash} retains the larger advantage.

Two limitations qualify this analysis. First, \textsc{PTC}'s confidence intervals include zero in most bins, limiting evidence for a consistent score gain. Second, \textsc{Tool-only} call count is an imperfect complexity proxy: high counts can reflect interface inefficiency rather than intrinsic task difficulty.

\subsection{Bash Efficiency}
\label{sec:results_bash_efficiency}

\begin{table}[!h]
\centering
\small
\setlength{\tabcolsep}{1.5pt}
\begin{tabularx}{\columnwidth}{@{}>{\raggedright\arraybackslash}X r r r r r r@{}}
\toprule
\textbf{Interface} & \shortstack[r]{\textbf{Cmds}\\\textbf{/call}} & \textbf{LOC} & \textbf{Composed} & \textbf{Loop} & \textbf{Failed} & \textbf{Repair} \\
\midrule
\multicolumn{7}{@{}l}{\emph{TheAgentCompany}} \\
\addlinespace[1pt]
\textsc{Bash} & 4.0 & 2 & 94.3\% & 27.5\% & 7.9\% & 76.3\% \\
\textsc{Bash+\allowbreak Tool} & 3.0 & 2 & 93.9\% & 25.7\% & 10.8\% & 58.7\% \\
\textsc{Bash+\allowbreak Synthesis} & 3.0 & 1 & 88.0\% & 24.0\% & 7.9\% & 70.0\% \\
\midrule
\multicolumn{7}{@{}l}{\emph{APEX-Agents}} \\
\addlinespace[1pt]
\textsc{Bash} & 1.5 & 12 & 96.5\% & 67.8\% & 4.4\% & 55.8\% \\
\textsc{Bash+\allowbreak Tool} & 1.5 & 16 & 97.2\% & 74.0\% & 9.3\% & 35.8\% \\
\textsc{Bash+\allowbreak Synthesis} & 2.0 & 5 & 93.3\% & 48.2\% & 5.4\% & 51.1\% \\
\bottomrule
\end{tabularx}
\caption{Shell statistics. \emph{Cmds/call}: median commands. \emph{LOC}: lines of code. \emph{Composed}: calls with pipe, logical-operator, substitution, control-flow, heredoc, multiline, or nontrivial-redirect markers; \emph{loop}: control-flow markers. \emph{Failed}: failures. \emph{Repair}: failures whose next call shares the first command.}
\label{tab:bash_efficiency}
\end{table}

\begin{figure*}[!t]
\centering
\includegraphics[width=0.95\textwidth]{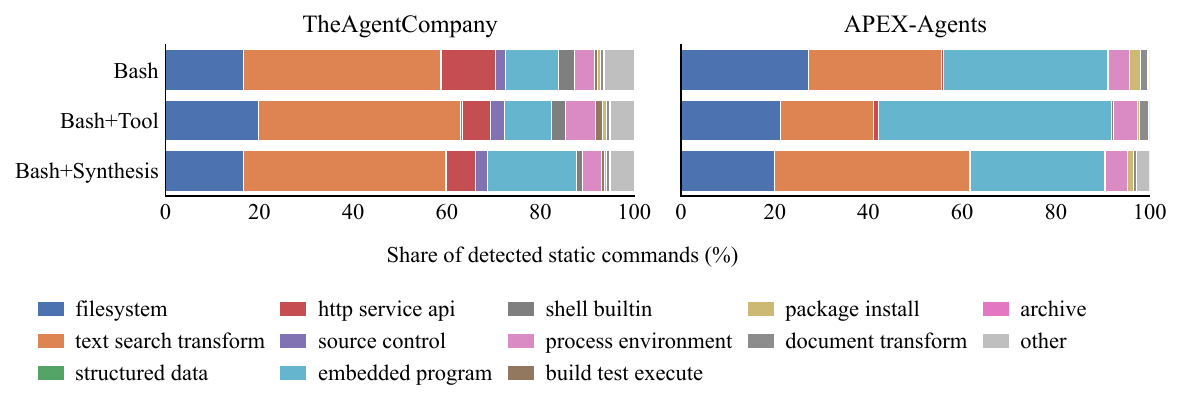}
\caption{Static shell-command shares by executable family, pooled across models (Table~\ref{tab:bash_families_glossary}). Bars: 12 families plus \emph{other}, summing to $100\%$ per benchmark/interface.}
\label{fig:bash_families}
\end{figure*}

We examine bash efficiency via composition, code length, failures, and immediate same-command retries (Table~\ref{tab:bash_efficiency}). Composition remains the norm across all three shell interfaces, even when typed or synthesized tools are available.

\textsc{Bash+Tool} reduces shell use without improving the reliability of remaining shell calls. It combines the lowest median shell calls per shell-using task with the highest failure rate and lowest retry rate in both benchmarks (Appendix~\ref{app:bash_efficiency_detail}).

\textsc{Bash+Synthesis} packages code behind shorter invocations, reflected in its lowest median lines of code (LOC) and composition fraction on both benchmarks. Calls invoking synthesized tools are also shorter on average than other synthesis calls, supporting code packaging as a practical benefit (Appendix~\ref{app:bash_efficiency_detail}).

Benchmark differences reveal two uses of the shell. TheAgentCompany combines more static commands into short calls, with text utilities and HTTP clients more prominent (Figure~\ref{fig:bash_families}). APEX-Agents uses longer embedded programs; most control-flow flags disappear when embedded and quoted bodies are stripped. Shell access can therefore support both utility-based workflows and file analysis.

\subsection{Tool Synthesis}
\label{sec:results_tool_synthesis}

We examine whether synthesized tools are reused reliably and whether their use is associated with better quality or lower token use. Agents invoke most tools they create, but fewer reach later tasks (Table~\ref{tab:synthesis_volume}). On APEX-Agents, GPT-5.5 creates fewer tools than Opus-4.8 yet uses all in later tasks. Appendix~\ref{app:tool_synthesis_detail} details the tools and measurements.

Reuse does not consistently improve call success. Reused tools and fresh inline code have similar success rates with Opus-4.8, but reused tools fare worse with GPT-5.5.

Tasks that invoke synthesized tools show no detectable mean score gain over \textsc{Bash} on either benchmark (Table~\ref{tab:synthesis_attribution}). On APEX-Agents, token differences relative to \textsc{Bash} are larger in the invoked subset than in the non-invoked subset.

\begin{table}[!h]
\centering
\small
\setlength{\tabcolsep}{1.5pt}
\begin{tabular}{@{}l r r r r r@{}}
\toprule
\textbf{Model} & \textbf{Created} & \textbf{Invoked} & \shortstack[r]{\textbf{Cross-task}\\\textbf{reuse}} & \shortstack[r]{\textbf{Reuse}\\\textbf{success}} & \shortstack[r]{\textbf{Fresh}\\\textbf{success}} \\
\midrule
\multicolumn{6}{@{}l}{\emph{TheAgentCompany}} \\
Opus-4.8 & 21 & 19 (90.5\%) & 12 (57.1\%) & 94.0\% & 95.3\% \\
GPT-5.5 & 12 & 10 (83.3\%) & 6 (50.0\%) & 83.7\% & 88.3\% \\
\midrule
\multicolumn{6}{@{}l}{\emph{APEX-Agents}} \\
Opus-4.8 & 32 & 29 (90.6\%) & 14 (43.8\%) & 96.4\% & 96.1\% \\
GPT-5.5 & 10 & 10 (100.0\%) & 10 (100.0\%) & 80.1\% & 86.2\% \\
\addlinespace[2pt]
\bottomrule
\end{tabular}
\caption{Tool synthesis reuse statistics and success rates. \emph{Created}: distinct agent-authored root Python paths, excluding seeds, initializers, and dependencies. \emph{Invoked}: tools used from creation onward; \emph{cross-task reuse}: used in later tasks. Percentages divide by created tools. \emph{Reuse success}: successful calls of synthesized tools; \emph{fresh success}: successful calls of fresh shell commands.}
\label{tab:synthesis_volume}
\end{table}

\begin{table}[!h]
\centering
\footnotesize
\setlength{\tabcolsep}{2pt}
\begin{tabular}{@{}l r r r r r@{}}
\toprule
\textbf{Subset} & \textbf{$N$} & \shortstack[r]{\textbf{Median}\\\textbf{$\Delta$ score}} & \shortstack[r]{\textbf{Mean}\\\textbf{$\Delta$ score}} & \shortstack[r]{\textbf{Median}\\\textbf{$\Delta$ tokens}} & \shortstack[r]{\textbf{Mean}\\\textbf{$\Delta$ tokens}} \\
\midrule
\multicolumn{6}{@{}l}{\emph{TheAgentCompany}} \\
Invoked & 228 & +0.0\,pp & -0.6\,pp & +0.6k & +1.1k \\
Not invoked & 120 & +0.0\,pp & -3.7\,pp & +0.8k & +3.3k \\
\midrule
\multicolumn{6}{@{}l}{\emph{APEX-Agents}} \\
Invoked & 559 & +0.0\,pp & +0.1\,pp & +31.4k & +63.0k \\
Not invoked & 401 & +0.0\,pp & -0.6\,pp & +8.5k & -0.3k \\
\addlinespace[2pt]
\bottomrule
\end{tabular}
\caption{Paired differences of \textsc{Bash+Synthesis} minus \textsc{Bash}, pooled across models. \emph{Subset}: whether synthesized tools are invoked; $N$: number of tasks in the corresponding subset. $\Delta$ score: task-score difference; $\Delta$ tokens: total-token difference.}
\label{tab:synthesis_attribution}
\end{table}

\subsection{Programmatic Tool Calling}
\label{sec:results_ptc}

We compare \textsc{PTC} with \textsc{Tool-only} to assess call batching and changes in agent-visible invocations and token use (Table~\ref{tab:ptc_composition}). Despite modest batching and invocation reductions, \textsc{PTC} substantially reduces uncached input and cache-read tokens on both benchmarks, with broadly similar output-token use. These savings may reflect processing intermediate results within programs and returning only selected outputs, reducing the context carried into subsequent model requests.

\begin{table}[!h]
\centering
\footnotesize
\setlength{\tabcolsep}{1.5pt}
\begin{tabular}{@{}l r r r r r@{}}
\toprule
\textbf{Benchmark} & \shortstack[r]{\textbf{Comp.}\\\textbf{ratio}} & \textbf{$\Delta$ call} & \shortstack[r]{\textbf{$\Delta$ uncached}\\\textbf{input}} & \shortstack[r]{\textbf{$\Delta$ cache-}\\\textbf{read}} & \shortstack[r]{\textbf{$\Delta$}\\\textbf{output}} \\
\midrule
TheAgentCompany & 2.0 & -4.0 & -97.6k & -86.8k & +0.2k \\
APEX-Agents & 2.0 & -3.0 & -45.4k & -44.0k & -0.4k \\
\bottomrule
\end{tabular}
\caption{Programmatic tool calling composition and token efficiency. \emph{Comp.\ ratio}: per-task typed calls per \textsc{PTC} execution. Deltas: \textsc{PTC} minus \textsc{Tool-only}, paired by task and model; tool calls count; uncached input tokens; cache-read tokens; output tokens. }
\label{tab:ptc_composition}
\end{table}

\section{Discussion}
\label{sec:discussion}
    Shell-only agents offer the strongest overall task performance and cost efficiency with the simplest tool scaffold in this study. Adding a typed catalog or persistent synthesized-tool library to the shell provides no consistent quality gain. Programmatic tool calling reduces token use over direct typed-tool calling with broadly similar or slightly better performance depending on workplace scenarios, while underperforming shell-only on both quality and cost. Practitioners could use shell-only agents where arbitrary execution can be isolated, and programmatic tool calling where security or compliance requirements restrict actions to an approved catalog.

Our results suggest starting with Opus-4.8 for shell-only coding and cross-application work, or GPT-5.5 for token-efficient direct tool calls. For document-heavy analysis and programmatic tool calling, practitioners could weigh Opus-4.8's higher quality against GPT-5.5's lower token use. As frontier models evolve, these recommendations may not carry over to later releases in either family.

Enterprise platforms could expose direct typed-tool calling, shell-only execution, and programmatic orchestration restricted to an approved catalog as configurable options. Practitioners could select among them according to task needs and whether general code execution is permitted.

Platforms could provide easy-to-navigate evaluation infrastructure that lets practitioners A/B-test interfaces and models on their own workplace scenarios. With permissions and human review held constant, teams could compare how often outputs meet users' needs, how much manual revision they require, operating cost, and operational failures. These evaluations could help practitioners choose the simplest tool interface that meets their workplace's quality, cost, and permission requirements.

\section*{Acknowledgments}
We thank the Microsoft Copilot Studio team for their thoughtful feedback and constructive discussions throughout this work. We are especially grateful to Soufiane Loukili for his sponsorship and continued support of this research.

\clearpage

\bibliography{aaai2027}

\appendix
\section{Harness and Prompts}
\label{app:method}

System prompts for all five tool interfaces share a benchmark-domain specific identity paragraph and benchmark task text, and varies in tool lists. Only synthesis and programmatic tool calling add interface-specific guidance (\S\ref{sec:method}). Synthesis encourages packaging reusable steps as parameterized commands but does not require tool creation.

The GitHub Copilot SDK supplies the agent loop, not its default toolset or system prompts, as they got overridden in our benchmark setups.

\section{Benchmark Details}
\label{app:exp_setup}

\subsection{TheAgentCompany}
\label{app:exp_tac}

TheAgentCompany combines cross-application workflows with local coding and single-application tasks (Figure~\ref{fig:environment_tac}). Examples include screening resumes from ownCloud against hiring requirements in RocketChat and synchronizing Plane issues with GitLab.

The typed catalog covers task-relevant service APIs and workspace operations, not every shell-accessible endpoint (Table~\ref{tab:tac_tool_catalogue}). With no canonical benchmark system prompt, we author a shared identity and vary only tools and interface-specific guidance (Appendix~\ref{app:method}).

Uniform coworker patches (\S\ref{sec:exp_tac}) recover missed channel messages on the next turn, prevent thread closure while a reply is expected, and reset Redis between tasks to prevent memory carryover. GPT-5 coworkers remain fixed across models and interfaces; scoring follows Table~\ref{tab:ablation_summary}.

During development, we observed reward hacking in shell-enabled interfaces: agents extracted answers from evaluator scripts in TheAgentCompany's public repository rather than completing the intended workflows, or read simulated-coworker files instead of contacting the coworkers. In all reported TheAgentCompany runs, we mitigated these shortcuts by blocking network access to that repository and encrypting the coworker information.

\begin{table*}[t]
\centering
\footnotesize
\begin{tabularx}{\textwidth}{@{}l >{\raggedright\arraybackslash}X@{}}
\toprule
\textbf{Service} & \textbf{Tools} \\
\midrule
Workspace & \texttt{tac\_workspace\_read\_file}, \texttt{tac\_workspace\_list\_dir}, \texttt{tac\_workspace\_write\_file}, \texttt{tac\_workspace\_write\_xlsx}, \texttt{tac\_workspace\_write\_docx}, \texttt{tac\_workspace\_write\_pptx} \\
\midrule
Rocket.Chat & \texttt{rocketchat\_search\_users}, \texttt{rocketchat\_list\_channels}, \texttt{rocketchat\_get\_channel\_history}, \texttt{rocketchat\_get\_direct\_history}, \texttt{rocketchat\_send\_channel\_message}, \texttt{rocketchat\_send\_direct\_message}, \texttt{rocketchat\_create\_channel}, \texttt{rocketchat\_update\_channel\_members}, \texttt{rocketchat\_set\_channel\_role} \\
\midrule
ownCloud & \texttt{owncloud\_search\_files}, \texttt{owncloud\_list\_directory}, \texttt{owncloud\_download\_file}, \texttt{owncloud\_read\_text}, \texttt{owncloud\_read\_pdf}, \texttt{owncloud\_read\_xlsx}, \texttt{owncloud\_read\_docx}, \texttt{owncloud\_read\_pptx}, \texttt{owncloud\_read\_csv}, \texttt{owncloud\_read\_image}, \texttt{owncloud\_upload\_file}, \texttt{owncloud\_create\_folder}, \texttt{owncloud\_move\_or\_copy\_file}, \texttt{owncloud\_create\_share\_link}, \texttt{owncloud\_delete\_file} \\
\midrule
GitLab & \texttt{gitlab\_search\_projects}, \texttt{gitlab\_list\_repository\_tree}, \texttt{gitlab\_get\_file}, \texttt{gitlab\_read\_files}, \texttt{gitlab\_search\_code}, \texttt{gitlab\_create\_branch}, \texttt{gitlab\_commit\_files}, \texttt{gitlab\_list\_issues}, \texttt{gitlab\_get\_issue}, \texttt{gitlab\_create\_issue}, \texttt{gitlab\_update\_issue}, \texttt{gitlab\_list\_merge\_requests}, \texttt{gitlab\_create\_merge\_request}, \texttt{gitlab\_list\_commits}, \texttt{gitlab\_list\_pipelines}, \texttt{gitlab\_create\_release}, \texttt{gitlab\_create\_or\_update\_wiki\_page}, \texttt{gitlab\_create\_project}, \texttt{gitlab\_update\_project\_settings} \\
\midrule
Plane & \texttt{plane\_list\_workspaces\_projects}, \texttt{plane\_create\_project}, \texttt{plane\_delete\_project}, \texttt{plane\_list\_members}, \texttt{plane\_list\_states}, \texttt{plane\_list\_cycles}, \texttt{plane\_search\_issues}, \texttt{plane\_get\_issue}, \texttt{plane\_create\_issue}, \texttt{plane\_update\_issue}, \texttt{plane\_get\_analytics\_summary} \\
\bottomrule
\end{tabularx}
\caption{TheAgentCompany's 60 typed tools by service: callable functions for \textsc{PTC}; JSON tools for \textsc{Tool-only} and \textsc{Bash+Tool}.}
\label{tab:tac_tool_catalogue}
\end{table*}

\subsection{APEX-Agents}
\label{app:exp_apex}

APEX-Agents spans 33 scenario worlds in investment banking, management consulting, and corporate law, with equal task counts across domains. Tasks include contract review, spreadsheet valuation, and survey reconciliation using shared files. Most request written answers; others require creating or editing documents, spreadsheets, or presentations.

The APEX-Agents snapshot combines the application servers in Table~\ref{tab:apex_tool_catalogue} with a Unix \texttt{sh -c} server. Shell execution enforces filesystem restrictions, environment-variable scrubbing, and execution limits rather than a Python-only restriction. Interface exposure follows \S\ref{sec:method}.

The catalog pins Archipelago revision on June 5, 2026, recording tool names, descriptions, JSON schemas, and the environment-image identifier. It predates the July 16 eleven-server revision's FMP and EDGAR SEC integrations, avoiding FMP's paid, external-API-key online mode.

The prompt adapts Archipelago's reference agent with short domain-specific instructions, but inference uses our harness. The Archipelago-derived judge evaluates each criterion from the task, final response, and workspace-file changes; score is the fraction satisfied, and passing requires all criteria.

Reasoning effort stays at the provider default across interfaces for each model. The public APEX leaderboard instead uses maximal reasoning and the latest server snapshot, so absolute scores are not directly comparable.

\begin{table*}[!t]
\centering
\footnotesize
\begin{tabularx}{\textwidth}{@{}l >{\raggedright\arraybackslash}X@{}}
\toprule
\textbf{Server} & \textbf{Tools} \\
\midrule
Calendar & \texttt{calendar}, \texttt{calendar\_schema} \\
\midrule
Chat & \texttt{chat}, \texttt{chat\_schema} \\
\midrule
Docs & \texttt{docs}, \texttt{docs\_schema} \\
\midrule
Filesystem & \texttt{get\_directory\_tree}, \texttt{get\_file\_metadata}, \texttt{list\_files}, \texttt{read\_image\_file}, \texttt{read\_text\_file}, \texttt{search\_files} \\
\midrule
Mail & \texttt{mail}, \texttt{mail\_schema} \\
\midrule
PDF & \texttt{pdf}, \texttt{pdf\_schema} \\
\midrule
Sheets & \texttt{sheets}, \texttt{sheets\_schema} \\
\midrule
Slides & \texttt{slides}, \texttt{slides\_schema} \\
\bottomrule
\end{tabularx}
\caption{APEX-Agents' 20 typed tools across 8 non-shell servers. Except filesystem, each pairs an action-dispatched tool with a schema tool.}
\label{tab:apex_tool_catalogue}
\end{table*}

\section{Additional Results}
\label{app:results}

\subsection{Scores by Domain}
\label{app:by_slice}

\begin{figure*}[t]
\centering
\includegraphics[width=0.9\textwidth]{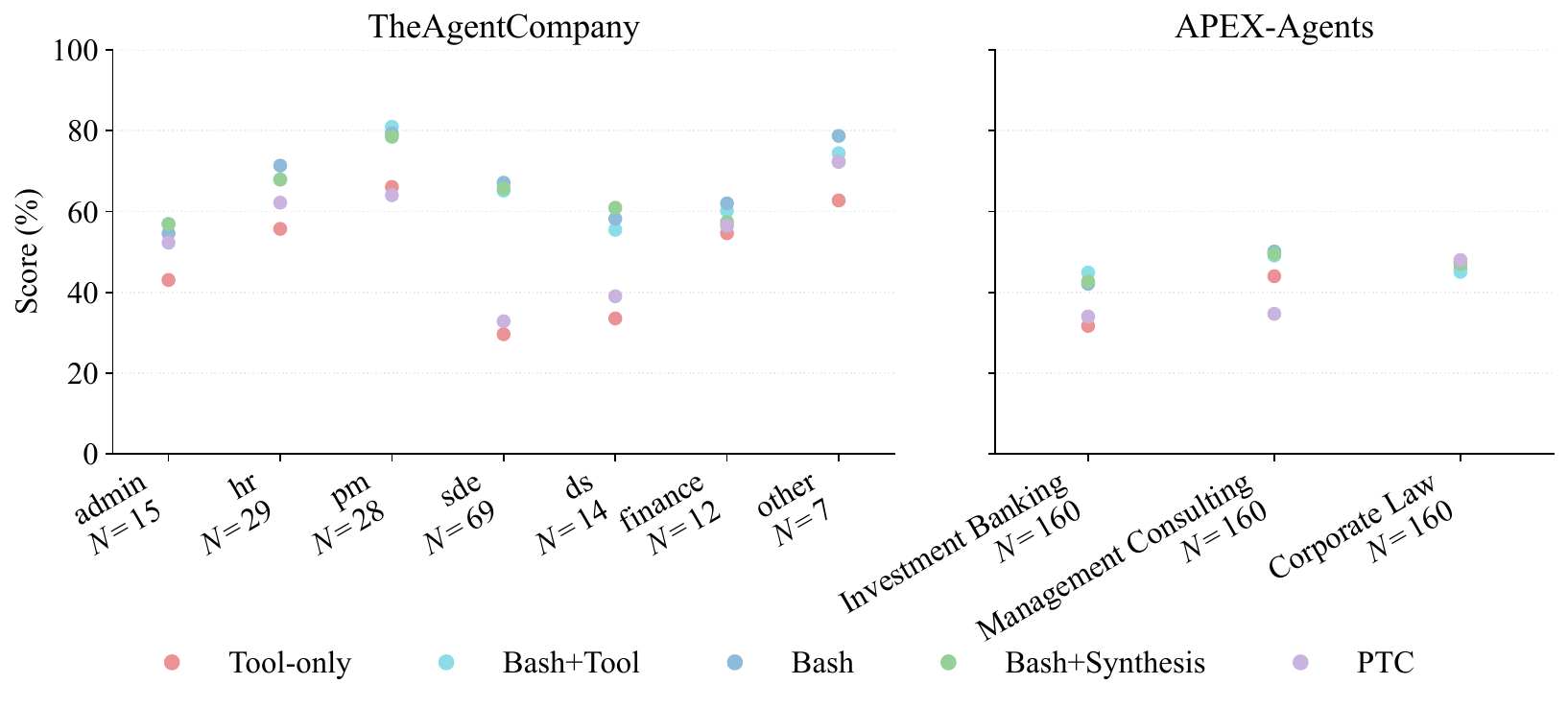}
\caption{Mean score (\%) across models by domain. Dots: interface/domain arms; colors: interfaces; $N$: tasks/domain. TheAgentCompany's \emph{other} pools \texttt{bm}, \texttt{ml}, \texttt{qa}, and \texttt{research}.}
\label{fig:by_slice}
\end{figure*}

We examine how differences between interfaces vary across enterprise domains or scenarios. The domain breakdown separates TheAgentCompany job families and APEX-Agents professional domains (Figure~\ref{fig:by_slice}). TheAgentCompany's smallest families are pooled using effort weighting in the \emph{other} domain.

Every domain follows the score ordering \textsc{Bash}~$\geq$~\textsc{Bash+Tool}~$\geq$~\textsc{Tool-only}, with the greatest separation in software engineering. Corporate Law shows the least sensitivity to interface choice, with all five interfaces closely grouped.

\subsection{Paired Score Differences}
\label{app:pairwise}

We examine how much task scores differ between interfaces and how uncertain these differences are. For each benchmark, we average interface score differences over matched task-model pairs from both models, giving each pair equal weight (Table~\ref{tab:pairwise}).

\begin{table*}[!t]
\centering
\small
\setlength{\tabcolsep}{4pt}
\begin{tabular}{@{}l r l c@{}}
\toprule
\textbf{Contrast (A vs.\ B)} & \textbf{Paired N} & \textbf{Mean score difference (pp, 95\% CI)} & \textbf{Wins / Ties / Losses} \\
\midrule
\multicolumn{4}{@{}l}{\emph{TheAgentCompany}} \\
\textsc{Bash+Tool} $-$ \textsc{Bash} & 348 & -0.6 {\scriptsize [-3.0, +1.9]} & 26 / 282 / 40 \\
\textsc{Bash+Tool} $-$ \textsc{Tool-only} & 348 & +24.1 {\scriptsize [+20.1, +28.3]} & 143 / 192 / 13 \\
\textsc{Bash} $-$ \textsc{Tool-only} & 348 & +24.6 {\scriptsize [+20.6, +29.0]} & 161 / 172 / 15 \\
\textsc{Bash+Synthesis} $-$ \textsc{Bash} & 348 & -1.6 {\scriptsize [-3.6, +0.3]} & 21 / 298 / 29 \\
\textsc{Bash} $-$ \textsc{PTC} & 348 & +21.2 {\scriptsize [+17.0, +25.5]} & 141 / 188 / 19 \\
\textsc{PTC} $-$ \textsc{Tool-only} & 348 & +3.5 {\scriptsize [+0.8, +6.2]} & 49 / 264 / 35 \\
\midrule
\multicolumn{4}{@{}l}{\emph{APEX-Agents}} \\
\textsc{Bash+Tool} $-$ \textsc{Bash} & 960 & -0.2 {\scriptsize [-2.4, +1.9]} & 149 / 642 / 169 \\
\textsc{Bash+Tool} $-$ \textsc{Tool-only} & 960 & +5.9 {\scriptsize [+3.6, +8.3]} & 208 / 620 / 132 \\
\textsc{Bash} $-$ \textsc{Tool-only} & 960 & +6.2 {\scriptsize [+3.6, +8.7]} & 220 / 596 / 144 \\
\textsc{Bash+Synthesis} $-$ \textsc{Bash} & 960 & -0.2 {\scriptsize [-2.4, +2.0]} & 164 / 638 / 158 \\
\textsc{Bash} $-$ \textsc{PTC} & 960 & +7.7 {\scriptsize [+5.2, +10.2]} & 228 / 589 / 143 \\
\textsc{PTC} $-$ \textsc{Tool-only} & 960 & -1.5 {\scriptsize [-3.9, +0.9]} & 174 / 596 / 190 \\
\bottomrule
\end{tabular}
\caption{\emph{Mean score difference}: mean paired A minus B in percentage points (positive favors A); brackets: 95\% bootstrap CIs. \emph{Paired N}: task-model pairs; \emph{wins / ties / losses}: pair counts summing to $N$.}
\label{tab:pairwise}
\end{table*}

\subsection{Shell Execution Details}
\label{app:bash_efficiency_detail}

We extend the Bash Efficiency analysis (\S\ref{sec:results_bash_efficiency}) with model-level shell patterns and limitations of the immediate-retry metric.

Shell syntax differs by model even within the same interface. On APEX-Agents, GPT-5.5 uses more heredoc markers for multiline input and fewer logical-chain markers (\texttt{\&\&}, \texttt{||}) than Opus-4.8 under \textsc{Bash}, \textsc{Bash+Tool}, and \textsc{Bash+Synthesis} (Table~\ref{tab:bash_efficiency_detail}).

The \emph{Repair} metric counts a retry only when the next shell call after a failure shares the first command (Table~\ref{tab:bash_efficiency}). It misses returns after intervening steps, which occur under \textsc{Bash+Tool} on both benchmarks, as in \texttt{python} $\rightarrow$ \texttt{pwd} $\rightarrow$ \texttt{python}. These sequences follow logged order and show a return to the executable, not a confirmed repair or response to the failure.

\begin{table*}[!t]
\centering
\small
\setlength{\tabcolsep}{4pt}
\begin{tabular}{@{}l l r r r r r r r r@{}}
\toprule
\textbf{Interface} & \textbf{Model} & \textbf{Calls/task} & \textbf{Cmds/call} & \textbf{Composed} & \textbf{Pipe} & \textbf{And/Or} & \textbf{Loop} & \textbf{Heredoc} & \textbf{Multi-line} \\
\midrule
\multicolumn{10}{@{}l}{\emph{TheAgentCompany}} \\
\addlinespace[1pt]
\textsc{Bash} & Opus-4.8 & 11 {\scriptsize [7, 16]} & 4.0 {\scriptsize [3, 5]} & 91.4\% & 56.4\% & 51.6\% & 23.6\% & 9.9\% & 51.1\% \\
 & GPT-5.5 & 14 {\scriptsize [9, 22]} & 4.0 {\scriptsize [2, 6]} & 97.3\% & 51.8\% & 50.3\% & 31.5\% & 33.3\% & 49.8\% \\
\addlinespace[2pt]
\textsc{Bash+Tool} & Opus-4.8 & 6 {\scriptsize [3, 11]} & 3.0 {\scriptsize [2, 5]} & 92.0\% & 47.9\% & 71.4\% & 19.3\% & 7.4\% & 31.3\% \\
 & GPT-5.5 & 6 {\scriptsize [3, 13]} & 3.0 {\scriptsize [2, 5]} & 95.7\% & 37.5\% & 57.5\% & 32.2\% & 32.3\% & 51.6\% \\
\addlinespace[2pt]
\textsc{Bash+Synthesis} & Opus-4.8 & 10 {\scriptsize [7, 13]} & 3.0 {\scriptsize [2, 4]} & 84.0\% & 39.9\% & 47.1\% & 23.8\% & 11.9\% & 38.7\% \\
 & GPT-5.5 & 18 {\scriptsize [12, 27]} & 3.0 {\scriptsize [2, 5]} & 92.1\% & 41.8\% & 59.8\% & 24.2\% & 27.5\% & 36.2\% \\
\addlinespace[2pt]
\midrule
\multicolumn{10}{@{}l}{\emph{APEX-Agents}} \\
\addlinespace[1pt]
\textsc{Bash} & Opus-4.8 & 8 {\scriptsize [5, 12]} & 2.0 {\scriptsize [2, 3]} & 93.9\% & 32.5\% & 48.6\% & 58.3\% & 16.8\% & 67.1\% \\
 & GPT-5.5 & 7 {\scriptsize [5, 11]} & 1.0 {\scriptsize [1, 2]} & 99.1\% & 38.0\% & 11.6\% & 77.3\% & 82.7\% & 83.8\% \\
\addlinespace[2pt]
\textsc{Bash+Tool} & Opus-4.8 & 3 {\scriptsize [1, 6]} & 2.0 {\scriptsize [1, 3]} & 94.7\% & 22.6\% & 34.2\% & 61.5\% & 34.2\% & 78.2\% \\
 & GPT-5.5 & 3 {\scriptsize [2, 6]} & 1.0 {\scriptsize [1, 1]} & 99.7\% & 23.1\% & 7.5\% & 86.5\% & 97.0\% & 97.2\% \\
\addlinespace[2pt]
\textsc{Bash+Synthesis} & Opus-4.8 & 10 {\scriptsize [6, 16]} & 2.0 {\scriptsize [2, 3]} & 88.8\% & 41.9\% & 28.9\% & 42.7\% & 14.8\% & 47.2\% \\
 & GPT-5.5 & 9 {\scriptsize [6, 12]} & 2.0 {\scriptsize [1, 5]} & 97.9\% & 56.4\% & 24.1\% & 53.6\% & 53.9\% & 78.9\% \\
\addlinespace[2pt]
\bottomrule
\end{tabular}
\caption{Bash metrics for recoverable-command calls by model/interface. \emph{Calls/task}: median per shell-using task; \emph{Cmds/call}: median heuristic static commands (minimum one), excluding embedded code and loop iterations. Brackets: 25th/75th percentiles. Other columns: nonexclusive call percentages per benchmark/model/interface, using source-text markers including embedded code. \emph{Composed}: any listed style, substitution, or redirection with multiple commands. \emph{Pipe}: \texttt{|}, not \texttt{||}; \emph{and/or}: \texttt{\&\&} or \texttt{||}; \emph{loop}: \texttt{for}, \texttt{while}, \texttt{until}, \texttt{if}, \texttt{case}, or \texttt{select}; \emph{heredoc}: \texttt{<<}; \emph{multi-line}: multiple submitted lines. Markers do not establish execution.}
\label{tab:bash_efficiency_detail}
\end{table*}

\begin{table*}[!t]
\centering
\small
\setlength{\tabcolsep}{6pt}
\begin{tabular}{@{}l p{0.72\linewidth}@{}}
\toprule
\textbf{Family} & \textbf{Executables} \\
\midrule
\emph{filesystem} & \texttt{ls}, \texttt{find}, \texttt{pwd}, \texttt{cd}, \texttt{stat}, \texttt{du}, \texttt{df}, \texttt{tree}, \texttt{mkdir}, \texttt{rm}, \texttt{mv}, \texttt{cp}, \texttt{touch}, \texttt{chmod}, \texttt{chown}, \texttt{ln}, \texttt{file}, \texttt{readlink}, \texttt{realpath}, \texttt{basename}, \texttt{dirname}, \texttt{mktemp} \\
\addlinespace[2pt]
\emph{text\_search\_transform} & \texttt{grep}, \texttt{rg}, \texttt{sed}, \texttt{awk}, \texttt{cut}, \texttt{sort}, \texttt{uniq}, \texttt{head}, \texttt{tail}, \texttt{tr}, \texttt{wc}, \texttt{xargs}, \texttt{cat}, \texttt{echo}, \texttt{printf}, \texttt{tee}, \texttt{nl}, \texttt{paste}, \texttt{fmt}, \texttt{fold}, \texttt{tac}, \texttt{rev}, \texttt{expand}, \texttt{unexpand}, \texttt{diff}, \texttt{cmp}, \texttt{patch}, \texttt{iconv} \\
\addlinespace[2pt]
\emph{structured\_data} & \texttt{jq}, \texttt{yq}, \texttt{sqlite3}, \texttt{duckdb}, \texttt{csvkit}, \texttt{xmllint}, \texttt{xmlstarlet} \\
\addlinespace[2pt]
\emph{http\_service\_api} & \texttt{curl}, \texttt{wget}, \texttt{http}, \texttt{httpie} \\
\addlinespace[2pt]
\emph{source\_control} & \texttt{git}, \texttt{gh}, \texttt{hg}, \texttt{svn} (mostly \texttt{git} in practice) \\
\addlinespace[2pt]
\emph{embedded\_program} & \texttt{python}, \texttt{python3}, \texttt{node}, \texttt{ruby}, \texttt{rscript}, \texttt{perl}, \texttt{php}, \texttt{bash}, \texttt{sh}, \texttt{lua}, \texttt{awk} (mostly \texttt{python}/\texttt{python3} in practice) \\
\addlinespace[2pt]
\emph{shell\_builtin} & \texttt{test}, \texttt{[}, \texttt{[[}, \texttt{read}, \texttt{set}, \texttt{unset}, \texttt{alias}, \texttt{unalias}, \texttt{shift}, \texttt{exit}, \texttt{return}, \texttt{local}, \texttt{declare}, \texttt{typeset}, \texttt{readonly}, \texttt{eval}, \texttt{exec}, \texttt{wait}, \texttt{trap}, \texttt{pushd}, \texttt{popd}, \texttt{dirs}, \texttt{help}, \texttt{let}, \texttt{expr}, \texttt{bc}, \texttt{dc} \\
\addlinespace[2pt]
\emph{process\_environment} & \texttt{ps}, \texttt{top}, \texttt{htop}, \texttt{kill}, \texttt{pkill}, \texttt{env}, \texttt{export}, \texttt{source}, \texttt{sleep}, \texttt{timeout}, \texttt{which}, \texttt{whereis}, \texttt{date}, \texttt{whoami}, \texttt{hostname}, \texttt{uname}, \texttt{id}, \texttt{groups}, \texttt{printenv}, \texttt{nohup}, \texttt{jobs}, \texttt{disown}, \texttt{true}, \texttt{false} \\
\addlinespace[2pt]
\emph{build\_test\_execute} & \texttt{pytest}, \texttt{npm}, \texttt{npx}, \texttt{yarn}, \texttt{pnpm}, \texttt{make}, \texttt{cmake}, \texttt{cargo}, \texttt{go}, \texttt{mvn}, \texttt{gradle}, \texttt{dotnet}, \texttt{poetry}, \texttt{uv} \\
\addlinespace[2pt]
\emph{package\_install} & \texttt{pip}, \texttt{pip3}, \texttt{apt}, \texttt{apt-get}, \texttt{dnf}, \texttt{yum}, \texttt{brew}, \texttt{choco} \\
\addlinespace[2pt]
\emph{archive} & \texttt{tar}, \texttt{gzip}, \texttt{gunzip}, \texttt{zip}, \texttt{unzip}, \texttt{xz}, \texttt{bzip2}, \texttt{bunzip2}, \texttt{7z}, \texttt{zstd} \\
\addlinespace[2pt]
\emph{document\_transform} & \texttt{pandoc}, \texttt{libreoffice}, \texttt{soffice}, \texttt{pdftotext}, \texttt{pdfinfo}, \texttt{tesseract}, \texttt{convert}, \texttt{magick} \\
\bottomrule
\end{tabular}
\caption{Static command families for Figure~\ref{fig:bash_families}: heuristic classification by executable; unmatched commands form \emph{other}.}
\label{tab:bash_families_glossary}
\end{table*}

\subsection{Synthesized Tools and Reuse}
\label{app:tool_synthesis_detail}

We extend the Tool Synthesis analysis (\S\ref{sec:results_tool_synthesis}) by examining tools without typed counterparts and the limits of reuse measurements.

Some synthesized tools have no direct typed-tool counterpart. APEX-Agents examples include financial-modeling scripts and general-purpose \texttt{extract\_text.py} (Table~\ref{tab:synthesis_tools}).

Reuse tracking captures observable tool history, not every tool an agent may create. It counts overwritten paths once and classifies a call as cross-task reuse only when the tool was created and invoked in an earlier task; write/update calls are excluded. Transient or deleted tools may be missed.

The success-rate comparison between cross-task tool reuse and fresh inline code is sensitive to argument rejections. Fresh inline code means the first occurrence of exact non-library inline code within a task. For GPT-5.5 on APEX-Agents, rejections are more frequent in reuse calls; excluding them leaves only a small success-rate difference. This sensitivity check retains unknown outcomes and measures whole-call status, not individual tool-body reliability.

\begin{table*}[t]
\centering
\small
\setlength{\tabcolsep}{5pt}
\begin{tabular}{@{}l r l@{}}
\toprule
\textbf{Script} & \textbf{Uses} & \textbf{Substitute} \\
\midrule
\multicolumn{3}{@{}l}{\emph{TheAgentCompany}} \\
\addlinespace[1pt]
\texttt{owncloud\_dav.py} & 122 & \textit{owncloud\_*} \\
\texttt{rocketchat\_cli.py} & 69 & \textit{rocketchat\_*} \\
\texttt{rocketchat\_dm.py} & 42 & \textit{rocketchat\_*} \\
\texttt{rc\_users.py} & 38 & \textit{rocketchat\_*} \\
\texttt{rocketchat\_read.py} & 24 & \textit{rocketchat\_*} \\
\texttt{rocketchat.py} & 19 & \textit{rocketchat\_*} \\
\texttt{plane\_api.py} & 16 & \textit{plane\_*} \\
\texttt{pdf\_text.py} & 11 & -- \\
\texttt{rc\_channel.py} & 10 & \textit{rocketchat\_*} \\
\texttt{pdf\_ocr.py} & 8 & -- \\
\texttt{gitlab\_api.py} & 5 & \textit{gitlab\_*} \\
\texttt{rc\_admin.py} & 5 & \textit{rocketchat\_*} \\
\texttt{make\_odt.py} & 3 & -- \\
\texttt{password\_validate.py} & 3 & -- \\
\texttt{pdf\_render.py} & 3 & -- \\
\midrule
\multicolumn{3}{@{}l}{\emph{APEX-Agents}} \\
\addlinespace[1pt]
\texttt{extract\_text.py} & 254 & -- \\
\texttt{office\_search.py} & 133 & -- \\
\texttt{pdftext.py} & 96 & \textit{pdf\_server\_*} \\
\texttt{xlsx\_search.py} & 80 & \textit{sheets\_server\_*} \\
\texttt{xlsx\_find.py} & 61 & \textit{sheets\_server\_*} \\
\texttt{docxtext.py} & 55 & \textit{docs\_server\_*} / \textit{filesystem\_server\_*} \\
\texttt{ocr\_pdf.py} & 53 & \textit{pdf\_server\_*} \\
\texttt{pdfgrep.py} & 41 & \textit{pdf\_server\_*} \\
\texttt{pptx\_text.py} & 29 & \textit{slides\_server\_*} \\
\texttt{xlsx\_dump.py} & 26 & \textit{sheets\_server\_*} \\
\texttt{rapidocr\_pdf.py} & 24 & \textit{pdf\_server\_*} \\
\texttt{xlsx\_recalc.py} & 23 & \textit{sheets\_server\_*} \\
\texttt{xlsx\_update\_recalc\_read.py} & 17 & \textit{docs\_server\_*} / \textit{filesystem\_server\_*} \\
\texttt{txt2docx.py} & 13 & \textit{docs\_server\_*} / \textit{filesystem\_server\_*} \\
\texttt{xlsx\_lo\_recalc.py} & 10 & \textit{sheets\_server\_*} \\
\bottomrule
\end{tabular}
\caption{Top \textsc{Bash+Synthesis} tools ranked by usage. \emph{Uses}: distinct invoking task-model pairs, pooled across models, not raw calls. \emph{Substitute}: closest typed-tool surface (heuristic); dash: no direct counterpart.}
\label{tab:synthesis_tools}
\end{table*}

\subsection{Failure Types}
\label{app:failure_taxonomy}

\begin{figure*}[!t]
\centering
\includegraphics[width=0.95\textwidth]{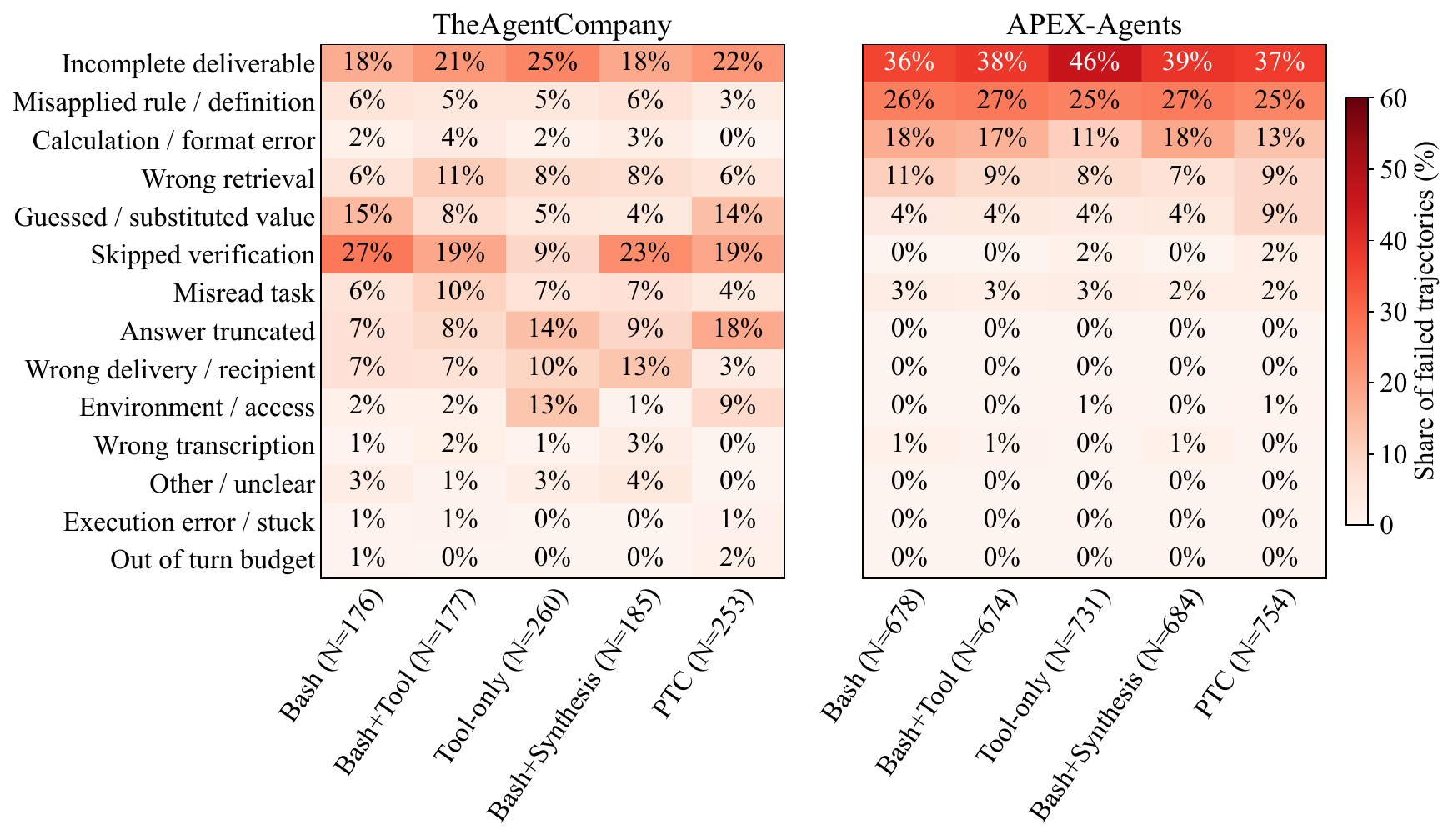}
\caption{Failure shares among trajectories with score $<1$, pooled across models. Rows: 14 root-cause categories ordered by cross-benchmark frequency; columns: benchmark/interface percentages summing to $100\%$; $N$: failed trajectories/arm.}
\label{fig:failure_taxonomy}
\end{figure*}

We examine how failure types vary across benchmarks and interfaces. An LLM judge assigns each failed trajectory to one failure category (Figure~\ref{fig:failure_taxonomy}).

Failure profiles differ more by benchmark than interface. APEX-Agents failures center on omitted requirements, misapplied domain rules or definitions, and incorrect numbers or formats across all five interfaces. TheAgentCompany spans incomplete deliverables, skipped verification, guessed values, wrong retrievals or deliveries, and environment/access errors.

TheAgentCompany's dominant failures vary by interface. \textsc{Tool-only} fails most often, chiefly through incomplete deliverables or environment/access errors. Shell-based interfaces most often declare completion without verification, while \textsc{PTC} has the highest share of final-message truncation.

\end{document}